\documentclass[11pt]{amsart}
\date{\today}

\usepackage[margin=1in]{geometry}
\usepackage{amsmath}
\usepackage{amsfonts,amssymb}
\newcommand{\R}{\mathbb{R}}

\numberwithin{equation}{section}

\newtheorem{thm}{Theorem}[section]

\newtheorem{lem}[thm]{Lemma}

\newcommand{\be}{\begin{equation}}
\newcommand{\ee}{\end{equation}}
\newcommand{\A}{\mathcal{A}}

\renewcommand{\epsilon}{\varepsilon}

\begin{document}
\title[Propagation of Correlations in Quantum Lattice Systems]{Propagation of Correlations\\[5pt]
in Quantum Lattice Systems}
\author{Bruno Nachtergaele}
\address{Department of Mathematics\\ University of California at Davis\\
Davis CA 95616, USA}
\email{bxn@math.ucdavis.edu} 
\author{Yoshiko Ogata}
\address{Department of Mathematics\\ University of California at Davis\\
Davis CA 95616, USA\\
and Department of Mathematical Sciences\\
University of Tokyo\\
Komaba, Tokyo, 153-8914 Japan}
\email{ogata@math.ucdavis.edu}
\author{Robert Sims}
\address{Department of Mathematics\\ University of California at Davis\\
Davis CA 95616, USA}
\email{rjsims@math.ucdavis.edu}

\begin{abstract}
We provide a simple proof of the Lieb-Robinson bound and use it to prove
the existence of the dynamics for interactions with polynomial decay.
We then use our results to demonstrate that there is an
upper bound on the rate at which correlations between
observables with separated support can accumulate as a 
consequence of the dynamics.
\end{abstract}
\maketitle 
\renewcommand{\thefootnote}{$ $}
\footnotetext{Copyright \copyright\ 2006 by the authors. This article may be
reproduced in its entirety for non-commercial purposes.}
\setcounter{section}{0}

%

\section{Introduction}

Recently, there has been increasing interest in understanding 
correlations in quantum lattice systems prompted by applications
in quantum information theory and computation 
\cite{schuch2005,cramer2005,bravyi2006,eisert2006}
and the study of complex networks \cite{hastings2004b}.
The questions that arise in the context of quantum information
and computation are sufficiently close to typical problems
in statistical mechanics that the methods developed in one
framework are often relevant in the other.
The bound on
the group velocity in quantum spin dynamics generated by a 
short-range Hamiltonian, which was proved by Lieb and Robinson 
more than three decades ago \cite{lieb1972}, is a case in point.
For example, as explained in \cite{bravyi2006}, the Lieb-Robinson bound
provides an upper bound on the speed of information transmission
through channels modeled by a quantum lattice systems with short-range
interactions.

The Lieb-Robinson bound plays a crucial role in the derivation
of several recent results. For some of these results it was useful, 
indeed necessary, to generalize and sharpen these bounds. Several
such improvements have recently appeared \cite{nachtergaele2005,hastings2005}. 
In this paper we provide a new proof of the Lieb-Robinson
bound (Theorem \ref{thm:lr}) and other estimates based on a norm-preserving 
property of the dynamics (see Lemma \ref{lem:normp}).
We apply this result to give upper bounds on the rate
at which correlations can be established between two separated
regions in the lattice for a general class of models (Theorem
\ref{thm:mare}). Moreover,
our bounds allow us to prove the existence of
the dynamics (Theorem \ref{thm:existence}), in the sense of a strongly 
continuous group of
automorphisms on the algebra of quasi-local observables for a 
larger class of interactions than was previously known
\cite{bratteli1997,simon1993,matsui1993}.

%
%

\subsection{The Set Up} We will be considering quantum spins systems defined over a set of
vertices $\Lambda$ equipped with a metric $d$. A finite dimensional 
Hilbert space $\mathcal{H}_x$ is assigned to each vertex $x \in
\Lambda$. In the most 
common cases $\Lambda$ is a graph, and the metric is given by the graph 
distance, $d(x,y)$, which may be the length of the shortest 
path of edges connecting $x$ and $y$ in the graph. 

For any finite subset $X \subset \Lambda$, the Hilbert space associated
with $X$ is the tensor product $\mathcal{H}_X = \bigotimes_{x \in X}\mathcal{H}_x$, and 
the set of corresponding observables supported in $X$ is denoted by
$\mathcal{A}_X=\mathcal{B}(\mathcal{H}_X)$, the bounded linear operators over $\mathcal{H}_X$. These local
observables form an algebra, and with the natural embedding of
$\mathcal{A}_{X_1}$ in $\mathcal{A}_{X_2}$ for any $X_1 \subset X_2$, one can define the
$C^*$-algebra of all observables, $\mathcal{A}$, as the norm completion of the
union of all local observable algebras $\mathcal{A}_{X}$ for finite $X \subset
\Lambda$. 

An interaction is a map $\Phi$ from the set of subsets
of $\Lambda$ to $\mathcal{A}$ with the property that $\Phi(X) \in \mathcal{A}_X$ and $\Phi(X) = \Phi(X)^*$
for all finite $X \subset \Lambda$. A quantum spin model is then defined to be the Hamiltonian,
expressed in terms of its interaction, given by
\begin{equation} \label{eq:defgenham}
H_{\Phi} := \sum_{ X \subset \Lambda} \Phi(X).
\end{equation}
For notational convenience, we will often drop the dependence of
$H_{\Phi}$ on $\Phi$. 

The dynamics, or time evolution, of a quantum spin model is 
the one-parameter group of automorphisms, $\{\tau_t\}_{t\in\R}$, defined by
\begin{equation}
\tau_t(A)=e^{itH} A e^{-itH}, \quad A \in \mathcal{A},
\end{equation}
which is always well defined for finite sets $\Lambda$. In the context of 
infinite systems, a boundedness condition on the interaction is
required in order for the finite-volume dynamics to converge to 
a strongly continuous one-parameter group of automorphisms on
$\mathcal{A}$.

To describe the interactions we wish to consider in this article, we
first put a condition on the set $\Lambda$; which is only relevant in
the event that $\Lambda$ is infinite. We assume that there 
exists a non-increasing function $F: [0, \infty) \to (0, \infty)$ 
for which:

\noindent i) $F$ is uniformly integrable over $\Lambda$, i.e.,
\begin{equation} \label{eq:fint}
\| \, F \, \| \, := \, \sup_{x \in \Lambda} \sum_{y \in \Lambda}
F(d(x,y)) \, < \, \infty,
\end{equation}

\noindent and 

\vspace{.3cm}

\noindent ii) $F$ satisfies
\begin{equation} \label{eq:intlat}
C \, := \, \sup_{x,y \in \Lambda} \sum_{z \in \Lambda} \frac{F \left( d(x,z) \right) \, F \left( d(z,y)
\right)}{F \left( d(x,y) \right)} \, < \, \infty.
\end{equation} 

Given a set $\Lambda$ equipped with a metric $d$, it is easy to
see that if $F$ satisfies i) and ii) above, then for any 
$a \geq 0$ the function
\begin{equation}
F_a(x) := e^{-ax} \, F(x),
\end{equation}
also satisfies i) and ii) with $\| F_a \| \leq \| F \|$ 
and $C_a \leq C$.

As a concrete example, take $\Lambda = \mathbb{Z}^d$ and $d(x,y) =
|x-y|$. In this case, one may take the function $F(x) = (1+x)^{-d - \epsilon}$ for any $\epsilon >0$. Clearly,
(\ref{eq:fint}) is satisfied, and a short calculation demonstrates
that (\ref{eq:intlat}) holds with 
\begin{equation}
C \, \leq \, 2^{d + \epsilon + 1} \, \sum_{n \in \mathbb{Z}^d}
\frac{1}{(1+|n|)^{d+ \epsilon}}.
\end{equation}
We also observe that, although the {\em purely exponential} function
$G(x) = e^{-ax}$, is integrable for $a>0$, i.e., it satisfies i), it does not
satisfy ii). This is evident from the fact that the cardinality of the
set $\{ z \in \mathbb{Z}^d : |x-z| +|z-y| -|x-y| = 0 \}$ is
proportional to $|x-y|$, and therefore, there exists no constant $C$
uniform in $|x-y|$.

To any set $\Lambda$ for which there exists a function $F$ satisfying
i) and ii) above, we define the set $\mathcal{B}_a(
\Lambda)$ to be those interactions $\Phi$ on $\Lambda$ which satisfy
\begin{equation} \label{eq:defnphia}
\| \Phi \|_a \, := \, \sup_{x,y \in \Lambda}  \sum_{X \ni x,y} \frac{ \| \Phi(X) \|}{F_a \left(
    d(x,y) \right)} \, < \, \infty.
\end{equation}

%

\section{Lieb-Robinson Estimates and Existence the Dynamics} \label{sec:lr}

%
%

\subsection{Lieb-Robinson Bounds}
We first present a variant of the Lieb-Robinson result which was 
first proven in \cite{nachtergaele2005,hastings2005}.

\begin{thm}[Lieb-Robinson Bound]\label{thm:lr}
Let $a \geq 0$ and take $\Lambda_1 \subset \Lambda$ a
finite subset. Denote by $\tau_t^{\Lambda_1}$ the time evolution corresponding
to a Hamiltonian 
\begin{equation}
H := \sum_{X \subset \Lambda_1} \Phi(X)
\end{equation} 
defined in terms of an interaction $\Phi \in \mathcal{B}_a(\Lambda)$. 
There exists a function $g: \R \to [0, \infty)$ with the property
that, given any pair of local observable $A \in
\mathcal{A}_{X}$ and $B \in \A_Y$ with $X, Y \subset
\Lambda_1$, one may estimate  
\begin{equation} \label{eq:lrbd1}
\left\| [ \tau_t^{\Lambda_1}(A), B ] \right\| \, \leq \, \frac{2 \, \| A \|
\, \|B \|}{C_a} \, g_a(t) \, \sum_{x \in X} \sum_{y \in
  Y} F_a \left( d(x,y) \right),
\end{equation}
for any $t \in \mathbb{R}$. Here the function
\begin{equation} \label{eq:gatt}
g_a(t) \, = \, \left\{ \begin{array}{cc} 
\left(e^{2 \, \| \Phi \|_a \, C_a \, |t|} - 1 \right)  & \mbox{ if }
d(X,Y)>0, \\  e^{2 \, \| \Phi \|_a \, C_a \, |t|} & \mbox{
  otherwise.} \end{array} \right.
\end{equation}
\end{thm}

\begin{proof}
Consider the function $f: \R \to \A$ defined by
\begin{equation}
f(t) := [ \tau_t^{\Lambda_1}(A), B].
\end{equation}
Clearly, $f$ satisfies the following differential equation
\begin{equation} \label{eq:lrde}
f'(t) = i \left[ f(t), \tau_t^{\Lambda_1} \left(H_X \right) \right] + i \left[
  \tau_t^{\Lambda_1}(A), \left[ \tau_t^{\Lambda_1}(H_X),B \right] \right],
\end{equation}
where we have used the notation
\begin{equation}
H_Y = \sum_{ \stackrel{Z \subset \Lambda_1:}{Z \cap Y \neq \emptyset}} \Phi(Z),
\end{equation}
for any subset $Y \subset \Lambda_1$. The first term in (\ref{eq:lrde}) above
is norm-preserving, and therefore the inequality
\begin{equation} \label{eq:lrineq1}
\| \, [ \tau_t^{\Lambda_1}(A), B] \, \|  \, \leq \, \| [A,B] \| \, + \, 2 \| A \| \, \int_0^{|t|} \,
\| \, [ \tau_s^{\Lambda_1}(H_X), B] \, \| \, ds  
\end{equation}
follows immediately from Lemma~\ref{lem:normp} and the automorphism property of
$\tau_t^{\Lambda_1}$. If we further define the quantity
\begin{equation}
C_B(X,t) := \sup_{A \in \A_X} \frac{ \| [ \tau_t^{\Lambda_1}(A), B ] \|}{ \|A \|},
\end{equation}
then (\ref{eq:lrineq1}) implies that 
\begin{equation} \label{eq:lrineq2}
C_B(X,t) \leq C_B(X,0) + 2 \sum_{ \stackrel{Z \subset \Lambda_1:}{Z \cap
    X \neq \emptyset}} \| \Phi(Z) \| \int_0^{|t|} C_B(Z, s) ds.
\end{equation}
Clearly, one has that
\begin{equation}
C_B(Z,0) \, \leq \, 2 \, \| B \| \, \delta_{Y}(Z), 
\end{equation}
where $\delta_{Y}(Z) = 0$ if $Z \cap Y = \emptyset$ and $\delta_{Y}(Z)
= 1$ otherwise. Using this fact, one may iterate (\ref{eq:lrineq2}) and find that 
\begin{equation}  \label{eq:seriesbd}
C_B(X,t) \, \leq \, 2 \| B \| \, \sum_{n=0}^{ \infty}
\frac{(2|t|)^n}{n!} a_n,
\end{equation}
where
\begin{equation}
a_n \, = \, \sum_{\stackrel{Z_1 \subset \Lambda_1:}{Z_1 \cap
    X \neq \emptyset}} \sum_{\stackrel{Z_2 \subset \Lambda_1:}{Z_2 \cap
    Z_1 \neq \emptyset}} \cdots \sum_{\stackrel{Z_n \subset \Lambda_1:}{Z_n \cap
    Z_{n-1} \neq \emptyset}} \prod_{i=1}^n \| \Phi(Z_i) \| \, \delta_Y(Z_n).
\end{equation}

For an interaction $\Phi \in \mathcal{B}_a(\Lambda)$, one may estimate that
\begin{equation} 
a_1 \, \leq \, \sum_{x \in X} \sum_{y \in Y} \sum_{Z \ni x,y} \|
\Phi(Z) \| \, \leq \, \| \Phi \|_a \, \sum_{x \in X} \sum_{y \in
  Y} F_a \left( d(x,y) \right).
\end{equation}
In addition,
\begin{eqnarray} 
a_2 & \leq & \sum_{x \in X} \sum_{y \in Y} \sum_{z \in \Lambda_1}
\sum_{ \stackrel{Z_1 \subset \Lambda_1:}{Z_1 \ni x,z}} 
\| \Phi(Z_1) \| \sum_{ \stackrel{Z_2 \subset \Lambda_1:}{Z_2 \ni z,y}}  \| \Phi(Z_2) \|  \nonumber \\ 
& \leq &   \| \Phi \|_a^2 \, \sum_{x \in X} \sum_{y \in Y} \sum_{z \in
  \Lambda} F_a \left( d(x,z) \right) \, F_a \left( d(z,y) \right)
\nonumber \\ & \leq & \| \Phi \|_a^2 \, C_a \, \sum_{x \in X} \sum_{y \in
  Y} F_a \left( d(x,y) \right),
\end{eqnarray}
using (\ref{eq:intlat}). With analogous arguments, one finds that
\begin{equation} \label{eq:aneq}
a_n \, \leq \, \| \Phi \|_a^n \, C_a^{n-1} \, \sum_{x \in X} \sum_{y \in
  Y} F_a \left( d(x,y) \right).
\end{equation}
Inserting (\ref{eq:aneq}) into (\ref{eq:seriesbd}) we see that
\begin{equation} \label{eq:lrbdd}
C_B(X,t) \leq \frac{2 \, \| B \| }{C_a} \, \mbox{exp} \left[2 \, \| \Phi
  \|_a \, C_a \, |t| \right] \sum_{x \in X} \sum_{y \in
  Y} F_a \left( d(x,y) \right),
\end{equation}
from which (\ref{eq:lrbd1}) immediately follows.

In the event that $d(X,Y)>0$, one has that $C_B(X,0) = 0$. 
For this reason the term corresponding to $a_0 = 0$, and therefore, 
the bound derived in (\ref{eq:lrbdd}) above holds with $e^{2 \| \Phi \|_a C_a |t|}$ replaced by 
$e^{2 \| \Phi \|_a C_a |t|}-1$. 
\end{proof}

We note that, for fixed local observables $A$ and $B$,
the bounds above are independent of the volume $\Lambda_1 \subset \Lambda$.

In the event that $\Phi \in \mathcal{B}_a(\Lambda)$ for some $a>0$, then the
bound in (\ref{eq:lrbd1}) implies that  
\begin{equation} \label{eq:vel}
\left\| [ \tau_t^{\Lambda_1}(A), B ] \right\| \, \leq \, \frac{2 \, \| A \|
\, \|B \|}{C_a} \, \| F \| \, \min(|X|,|Y|) \, e^{- a \,\left[
 d(X,Y) - \frac{2 \| \Phi \|_a C_a}{a} |t| \right]},
\end{equation}
which corresponds to a velocity of propagation given by
\begin{equation}
V_{\Phi} := \inf_{a>0} \frac{2 \| \Phi \|_a C_a}{a}.
\end{equation}
We further note that the bounds in (\ref{eq:lrbd1}) and 
(\ref{eq:vel}) above only require that one of
the observables have finite support; in particular, if  $|X|<
\infty$ and $d(X,Y)>0$, then the bounds are valid irrespective of the support of $B$.  

One can also view the Lieb-Robinson bound as a means of localizing the
dynamics. Let $\Lambda$ be finite and take $X \subset \Lambda$. 
Denote by $X^c = \Lambda \setminus X$. For any 
observable $A \in \A_{\Lambda}$ set
\begin{equation}
\langle A \rangle_{X^c} := \int_{\mathcal{U}(X^c)} U^* A U \, \mu(dU),
\end{equation}
where $\mathcal{U}(X^c)$ denotes the group of unitary operators over
the Hilbert space $\mathcal{H}_{X^c}$ and $\mu$ is the associated 
normalized Haar measure. 
It is easy to see that for any $A \in \A_{\Lambda}$, the quantity $\langle A
\rangle_{X^c} \in \A_X$ and the difference
\begin{equation} \label{eq:acomm}
\langle A \rangle_{X^c} \, - \, A \, = \, 
\int_{\mathcal{U}(X^c)} U^* \left[ A,  U  \right] \, \mu(dU).
\end{equation}
We can now combine these observations with the Lieb-Robinson bounds we
have proven. Let $A \in \A_X$ be a local observable, and choose
$\epsilon \geq 0$, $a>0$, and an interaction $\Phi \in
\mathcal{B}_a(\Lambda)$. We will denote by
\begin{equation}
B_t( \epsilon) \, = \, B(A, t, \epsilon) \,:= \,  \left\{ x \in \Lambda \, : \, d(x, X) \, \leq \,
  \frac{2 \| \Phi \|_a C_a}{a} \, |t| \, + \, \epsilon \, \right\},
\end{equation} 
the ball centered at $X$ with radius as specified above. 
For any $U \in \mathcal{U}(B_t^c(\epsilon))$, we clearly have that
\begin{equation}
d \left( X, {\rm supp}(U) \right) \, \geq \, \frac{2 \| \Phi \|_a
  C_a}{a} \,|t| \, + \epsilon,
\end{equation}
and therefore, using (\ref{eq:acomm}) above, we immediately
conclude that 
\begin{eqnarray}
\left\| \, \tau_t(A) \, - \, \langle \tau_t(A) \rangle_{B_t^c(\epsilon)} \, \right\| & 
\leq & \int_{\mathcal{U}(B_t^c(\epsilon))}  \left\| \,  \left[ \tau_t(A),  U
  \right] \, \right\| \, \mu(dU) \nonumber \\
& \leq & \frac{2 \, \| A \| \, |X| }{C_a} \, \| F \| \,
e^{- a \epsilon},
\end{eqnarray}
where for the final estimate we used (\ref{eq:vel}).

%
%

\subsection{Existence of the Dynamics}
As is demonstrated in \cite{bratteli1997}, one can use a Lieb-Robinson
bound to establish the existence of the dynamics for interactions
$\Phi \in \mathcal{B}_a( \Lambda)$. In the following we consider the thermodynamic limit
over a increasing exhausting sequence of finite subsets 
$\Lambda_n\subset\Lambda$.

\begin{thm}\label{thm:existence}
Let $a\geq 0$, and $\Phi \in \mathcal{B}_a(
  \Lambda)$. The dynamics $\{ \tau_t \}_{t \in \R}$ corresponding 
to $\Phi$ exists as a strongly continuous, one-parameter group of
automorphisms on $\A$. In particular,
\begin{equation}
\lim_{n\to \infty} \| \tau_t^{\Lambda_n}(A) - \tau_t(A) \| =
0
\end{equation}
for all $A \in \A$. The convergence is uniform for $t$ in compact sets
and independent of the choice of exhausting sequence $ \{ \Lambda_n \}$.
\end{thm}

\begin{proof}
Let $n>m$. Then, $\Lambda_m \subset \Lambda_n$. It is easy to verify that for 
any local observable $A \in \A_Y$,
\begin{equation}
\tau_t^{\Lambda_n}(A) - \tau_t^{\Lambda_m}(A) \, = \, \int_0^t \,
\frac{d}{ds} \left( \, \tau_s^{\Lambda_n} \tau_{t-s}^{\Lambda_m}(A) \, \right) \, ds, 
\end{equation} 
and therefore
\begin{equation} \label{eq:dynbd}
\left\| \tau_t^{\Lambda_n}(A) - \tau_t^{\Lambda_m}(A) \right\| \, \leq
\, \sum_{x \in \Lambda_n \setminus \Lambda_m} \sum_{X \ni x}
\int_0^{|t|} \left\| \, \left[ \, \Phi(X), \, \tau_s^{\Lambda_m}(A) \, \right] \, \right\| \, ds. 
\end{equation}
Applying Theorem~\ref{thm:lr}, we see that the right hand side of
(\ref{eq:dynbd}) is bounded from above by
\begin{equation}
2 \, \| A \| \, \int_0^{|t|} g_a(s) ds \, \sum_{x \in \Lambda_n \setminus \Lambda_m} \sum_{X \ni x}
 \| \Phi(X) \| \sum_{z \in X} \sum_{y \in Y} F_a \left( d(z,y) \right).
\end{equation}
Rewriting the sum on $X \ni x$ and $y \in X$ as the sum on
$y \in \Lambda$ and $X \ni x,y$, one finds that 
\begin{eqnarray}
\left\| \tau_t^{\Lambda_n}(A) - \tau_t^{\Lambda_m}(A) \right\| & \leq
& 2 \, \| A \| \, \| \Phi \|_a \, C_a \, \int_0^{|t|} g_a(s) ds \, \sum_{x
  \in \Lambda_n \setminus \Lambda_m} \sum_{z \in Y} F_a \left( d(x,z)
\right) \nonumber \\
& \leq & 2 \, \| A \| \, \| \Phi \|_a \, C_a \, \int_0^{|t|} g_a(s) ds \,
| Y | \, \sup_{z \in Y} \sum_{x
  \in \Lambda_n \setminus \Lambda_m} F_a \left( d(x,z) \right).
\end{eqnarray}
As $m,n\to \infty$, the above sum goes to zero. This proves that the sequence
is Cauchy and hence convergent. The remaining claims follow as in 
Theorem 6.2.11 of \cite{bratteli1997}.
\end{proof}

%
%
%

\section{Growth of Spatial Correlations}
The goal of this section is to prove Theorem~\ref{thm:mare} below
which bounds the rate at which correlations can accumulate, under the
influence of the dynamics, starting from a product state. 
\subsection{The Main Result}
Let $\Omega$ be a normalized product state, i.e.
$\Omega=\bigotimes_{x\in\Lambda}\Omega_x$, where for each $x$,
$\Omega_x$ is a state (not necessarily pure) for the systems 
at site $x$. We will denote by $ \langle \cdot \rangle$ the expectation with respect 
to $\Omega$, and prove 

\begin{thm}\label{thm:mare} Let $a \geq 0$, $\Phi \in \mathcal{B}_a(
  \Lambda)$, and take $\Omega$ to be a normalized product state as
  described above. Given $X, Y \subset \Lambda$ with $d(X,Y)>0$ and
  local observables $A \in \A_X$ and $B \in \A_Y$, one has that
\begin{equation} \label{eq:deccor}
\left| \, \langle \tau_t \left( A B \right)  \rangle \, -  \langle \tau_t(A) \rangle
  \, \langle \tau_t(B) \rangle \, \right| \, \leq \, 4 \, \| A \| \,
\| B \| \, \| F \| \, \left( \, |X| \, + \, |Y| \, \right) \, G_a(t)
\, e^{-a d(X,Y)},
\end{equation}
Here
\begin{equation}
G_a(t) \, = \, \frac{ C_a \, + \, \| F_a \|}{C_a} \, \| \Phi \|_a \,
\int_0^{|t|} g_a(s) \, ds,  
\end{equation}
and $g_a$ is the function which arises in the Lieb-Robinson estimate 
Theorem~\ref{thm:lr}.
\end{thm}

In the event that $a=0$, the bound above does not decay. However, the
estimate (\ref{eq:upbd}) below, which does decay, is valid. Moreover, 
a straight forward application of the techniques used below also provides
estimates on the increase of correlations, due to the dynamics,
for non-product states.

We begin by writing the interaction $\Phi$ as the sum of two terms,
one of which decouples the interactions between observables supported
near $X$ and $Y$.


%

\subsubsection{Decoupling the Interaction:}
Consider two separated local observables, i.e., $A \in \A_X$ and $B \in
\A_Y$ with $d(X,Y)>0$. Let
\begin{equation} \label{eq:sab}
S_{A,B} \, : = \, \left\{ y \in \Lambda \, : \, d(y,X) \, \leq \, \frac{d(X,Y)}{2} \, \right\},
\end{equation}
denote the ball centered at $X$ with distance $d(X,Y)/2$ from $Y$. For
any $\Phi \in \mathcal{B}_a (\Lambda)$, write
\begin{equation} \label{eq:intdec2}
\Phi \, = \, \Phi \left( 1 - \chi_{A,B} \right) \, + \, \Phi
\chi_{A,B} =: \Phi_1 + \Phi_2,
\end{equation}
where for any $Z \subset \Lambda$
\begin{equation} \label{eq:defchi}
\chi_{A,B}(Z) \, := \, \left\{ \begin{array}{cc} 1 & \mbox{if } Z \cap S_{A,B} \neq
    \emptyset \mbox{ and } Z \cap S_{A,B}^c \neq \emptyset, \\ 0 & \mbox{otherwise}. \end{array} \right.
\end{equation}
In this case, one has
\begin{lem} \label{lem:intlr}Let $a \geq 0$, $\Phi \in \mathcal{B}_a(\Lambda)$, and
consider any two separated local observables $A \in \A_X$
and $B \in \A_Y$ with $d(X,Y)>0$. Writing $\Phi = \Phi_1 + \Phi_2$,
as in (\ref{eq:intdec2}), one may show that
\begin{equation} \label{eq:declrbd}
\int_0^{|t|} \left\| \, \left[ \, H_2, \,
    \tau_s^{(1)}(O) \, \right] \, \right\| \, ds \, \leq \,
 2 \, \| O \| \, G_a(t) \, \sum_{o \in {\rm supp}(O)} \sum_{\stackrel{x \in \Lambda:}{2d(x,o) \geq d(X,Y)}} F_a
\left( d(x,o) \right), 
\end{equation}
is valid for observables $O \in \{ A,B \}$. One may take
\begin{equation}
G_a(t) \, = \, \frac{ C_a \, + \, \| F_a \|}{C_a} \, \| \Phi \|_a \,
\int_0^{|t|} g_a(s) \, ds,  
\end{equation}
where $g_a$ is the function from Theorem~\ref{thm:lr}.
\end{lem}
\begin{proof}
For $O \in \{ A, B \}$ and $s >0$,
\begin{equation}
\left\| \, \left[ \, H_2, \,
    \tau_s^{(1)}(O) \, \right] \, \right\| \, \leq  \,  \sum_{\stackrel{Z \subset \Lambda:}{Z \cap S_{A,B} \neq
    \emptyset, Z \cap S_{A,B}^c \neq \emptyset}}  \left\| \, \left[ \, \Phi(Z), \,
    \tau_s^{(1)}(O) \, \right] \, \right\|, 
\end{equation}
as is clear from the definition of $\chi_{A,B}$; see (\ref{eq:defchi}).
Applying Theorem~\ref{thm:lr} to each term above, we find that
\begin{equation}
\left\| \, \left[ \, \Phi(Z), \, \tau_s^{(1)}(O) \, \right] \,
\right\| \, \leq \, \frac{ 2 \, g_a(s) \, \| O \| \, \| \Phi(Z) \|}{C_a} \, \sum_{z \in Z} \sum_{o
  \in \mbox{supp}(O)} F_a \left( d(z,o) \right).
\end{equation} 
One may estimate the sums which appear above as follows:
\begin{eqnarray}
 \sum_{\stackrel{Z \subset \Lambda:}{Z \cap S_{A,B} \neq
    \emptyset, Z \cap S^c_{A,B} \neq \emptyset}} \sum_{z \in Z} & = & 
 \sum_{\stackrel{Z \subset \Lambda:}{Z \cap S_{A,B} \neq
    \emptyset, Z \cap S^c_{A,B} \neq \emptyset}} \left( 
\sum_{\stackrel{z \in Z:}{z \in S_{A,B}}} + 
\sum_{\stackrel{z \in Z:}{z \in S^c_{A,B}}} \right) \nonumber \\
& \leq & \sum_{z \in S_{A,B}} \sum_{x \in S^c_{A,B}} \sum_{Z \ni z,x}
+ \sum_{z \in S^c_{A,B}} \sum_{x \in S_{A,B}} \sum_{Z \ni z,x},  
\end{eqnarray}
and therefore, we have the bound
\begin{equation}
 \int_0^{|t|} \left\| \, \left[ \, H_2, \,
    \tau_s^{(1)}(O) \, \right] \, \right\| \, ds \, \leq \,
\frac{2 \| O \|}{C_a} \, \left( S_1 + S_2 \right) \, \int_0^{|t|} g_a(s) ds ,
\end{equation}
where
\begin{equation}
S_1 \, = \, \sum_{z \in S_{A,B}} \sum_{x \in S^c_{A,B}} \sum_{Z \ni
  z,x} \, \| \Phi(Z) \| \, \sum_{o
  \in \mbox{supp}(O)} F_a \left( d(z,o) \right) 
\end{equation}
and
\begin{equation}
S_2 \, = \,  \sum_{z \in S^c_{A,B}} \sum_{x \in S_{A,B}} \sum_{Z \ni
  z,x} \, \| \Phi(Z) \| \, \sum_{o
  \in \mbox{supp}(O)} F_a \left( d(z,o) \right).
\end{equation}

In the event that the observable $O=A$, then one may bound $S_1$ by
\begin{eqnarray}
S_1 & \leq &  \| \Phi \|_a \, \sum_{z \in S_{A,B}} \sum_{x \in
  S^c_{A,B}} F_a \left( d(z,x) \right) \, \sum_{y
  \in X} F_a \left( d(z,y) \right)   \\
& \leq &  C_a \, \| \Phi \|_a \, \sum_{x \in
  S^c_{A,B}} \sum_{y \in X} F_a \left( d(x,y) \right) \nonumber 
\end{eqnarray}
and similarly,
\begin{eqnarray}
S_2 & \leq &  \| \Phi \|_a \, \sum_{z \in S^c_{A,B}} \sum_{x \in
  S_{A,B}} F_a \left( d(z,x) \right) \, \sum_{y
  \in X} F_a \left( d(z,y) \right)  \\
& \leq &  \| F_a \| \, \| \Phi \|_a \, \sum_{z \in
  S^c_{A,B}} \sum_{y \in X} F_a \left( d(z,y) \right) \nonumber 
\end{eqnarray}
An analogous bound holds in the case that $O=B$. 
We have proven (\ref{eq:declrbd}).
\end{proof}

%
%

\subsubsection{Proof of Theorem~\ref{thm:mare}:}
To prove Theorem~\ref{thm:mare}, we will first provide an
estimate which measures the effect on the dynamics 
resulting from dropping certain interaction terms. 

\begin{lem} \label{lem:difham} Let $ \Phi_0 = \Phi_1 + \Phi_2$ be an interaction on
  $\Lambda$ for which each of the dynamics $\{ \tau_t^{(i)} \}_{t \in
    \R}$, for $i \in \{0,1,2 \}$, exists as a strongly continuous 
  group of $*$-automorphisms on $\A$. Let $\{ A_t \}_{ t \in \R}$ be a
  differentiable family of quasi-local observables on $\A$. 
  The estimate 
\begin{equation} \label{eq:dropint}
\| \, \tau_t^{(0)}(A_t) \, - \, \tau_t^{(1)}(A_t) \,  \| \, \leq
\, \int_0^{|t|} \, \left\| [ H_2 , \tau_s^{(1)}(A_s) ] \right\| \, + \,
\left\|  \tau_s^{(0)}( \partial_sA_s) - \tau_s^{(1)}( \partial_s A_s)
  \right\|
\, ds,
\end{equation}
holds for all $t \in \R$. Here, for each $i \in \{0,1,2\}$, we denote
by $H_i$ the Hamiltonian corresponding to $\Phi_i$.
\end{lem}

\begin{proof}
Define the function $f: \mathbb{R} \to \A$ by
\begin{equation}
f(t) \, := \, \tau_t^{(0)}(A_t) \, - \, \tau_t^{(1)}(A_t).
\end{equation}
A simple calculation shows that $f$ satisfies the following differential equation:
\begin{equation} \label{eq:fder}
f'(t) \, = \, i  \left[ H_0, f(t) \right] \, + \, i
\left[ H_2, \tau_t^{(1)}(A_t) \right] \, + \, \tau_t^{(0)}( \partial_t
A_t) - \tau_t^{(1)}( \partial_t A_t),
\end{equation}
subject to the boundary condition $f(0)=0$. The first term appearing
on the right hand side of (\ref{eq:fder}) above is norm preserving, and therefore, Lemma~\ref{lem:normp}
implies that 
\begin{equation}
\| \, f(t) \,  \| \, \leq
\, \int_0^{|t|} \, \left\| [ H_2, \tau_s^{(1)}(A_s) ] \right\| \, + \,
\left\|  \tau_s^{(0)}( \partial_sA_s) - \tau_s^{(1)}( \partial_s A_s)
  \right\| \, ds,
\end{equation}
as claimed.
\end{proof}

We will now prove Theorem~\ref{thm:mare}. Denote by $B_t := B - \langle
\tau_t(B) \rangle $, and observe that proving (\ref{eq:deccor}) is
equivalent to bounding $| \langle \tau_t(AB_t) \rangle |$. Write 
$\Phi \, = \, \Phi_1 \, + \, \Phi_2$, as is done in (\ref{eq:intdec2}).
One easily sees that $\Phi_1$ decouples $A$ from $B$, i.e., 
\begin{equation} \label{eq:fac}
\langle \, \tau_t^{(1)}(AB) \, \rangle \, = \, \langle \,
\tau_t^{(1)}(A) \, \rangle \, \langle \, \tau_t^{(1)}(B) \, \rangle.
\end{equation}
Here, again, we have denoted by $\tau_t^{(1)}$ the time evolution corresponding to 
$\Phi_1$. It is clear that 
\begin{eqnarray} \label{eq:corbd1}
\left| \langle \tau_t(AB_t) \rangle \right| & \leq &  \left| \langle
  \tau_t^{(1)}(AB_t) \rangle \right| \, + \, \left| \langle
  \tau_t(AB_t) \, - \, \tau_t^{(1)}(AB_t) \rangle \right| \\
& \leq & \| A \| \, \left\| \tau_t(B) - \tau_t^{(1)}(B) \right\| \, +
\, \left\| \tau_t(AB_t) - \tau_t^{(1)}(AB_t) \right\|. \nonumber
\end{eqnarray}
Moreover, the second term on the right hand side above can be further estimated by
\begin{equation} \label{eq:corbd2}
 \left\| \tau_t(AB_t) - \tau_t^{(1)}(AB_t) \right\| \, \leq \, 2 \| B
 \| \,  \left\| \tau_t(A) - \tau_t^{(1)}(A) \right\| \, + \, \| A \|
 \,  \left\| \tau_t(B_t) - \tau_t^{(1)}(B_t) \right\|.
\end{equation}
Applying Lemma~\ref{lem:difham} to the bounds we have found in (\ref{eq:corbd1}) and
(\ref{eq:corbd2}) yields
\begin{equation}
\left| \langle \tau_t(AB_t) \rangle \right| \, \leq \, 2 \| A \| \,
\int_0^{|t|} \left\| \left[ H_2, \tau_s^{(1)}(B) \right] \right\| \,
ds \, + \, 2 \| B \| \,
\int_0^{|t|} \left\| \left[ H_2, \tau_s^{(1)}(A) \right] \right\| \, ds.
\end{equation}
In fact, we are only using (\ref{eq:dropint}) in trivial situations
where the second term, i.e., $\tau_s( \partial_s A_s) - \tau_s^{(1)}(
\partial_s A_s)$ is identically zero. Finally, using Lemma~\ref{lem:intlr}, we
find an upper bound on $| \langle \tau_t(AB_t) \rangle |$ of the form
\begin{equation} \label{eq:upbd}
4 \, \| A \| \, \| B \| \, G_a(t) \left( \sum_{x \in X} \sum_{ \stackrel{y \in
    \Lambda:}{2d(x,y) \geq d(X,Y)}} F_a \left( d(x,y) \right) \, + \, 
\sum_{y \in Y} \sum_{ \stackrel{x \in \Lambda:}{2d(x,y) \geq d(X,Y)}} F_a \left( d(x,y) \right) \,\right).
\end{equation}
Theorem~\ref{thm:mare} readily follows from (\ref{eq:upbd}) above.

%
%
%
\setcounter{equation}{0}  
\renewcommand{\theequation}{\thesection.\arabic{equation}}                                                                                                  
\begin{appendix}
\section{}
                                                                                                         
In this appendix, we recall a basic lemma about the growth of the solutions 
of first order, inhomogeneous differential equations. 

Let $\mathcal{B}$ be a Banach space. For each $t \in \R$, 
let $A( t) : \mathcal{B} \to \mathcal{B}$ be a linear
operator, and denote by $X( t)$ the solution of the 
differential equation 
\begin{equation} \label{eq:fode}
\partial_{t} X( t) \, = \, A( t) \, X( t)
\end{equation}
with boundary condition $X(0) = x_0 \in \mathcal{B}$.We say that the
family of operators $A(t)$ is {\em norm-preserving} if for 
every $x_0 \in \mathcal{B}$, the mapping $\gamma_{t} :
\mathcal{B} \to \mathcal{B}$ which associates $x_0 \to X( t)$,
i.e., $\gamma_{t}(x_0) = X( t)$, satisfies
\begin{equation} \label{eq:normp}
\| \, \gamma_{t}(x_0) \, \| \, = \, \| \, x_0 \, \| \quad \mbox{for all } t \in
\R.
\end{equation}

Some obvious examples are the case where $\mathcal{B}$ is a Hilbert space 
and $A(t)$ is anti-hermitian for each $t$, or when $\mathcal{B}$ 
is an $*$-algebra of operators on a Hilbert space with a spectral norm and,
for each $t$, $A(t)$ is a derivation commuting with the $*$-operation.

\begin{lem} \label{lem:normp} Let $A( t)$, for $t \in \R$, be a family of 
norm preserving opeartors in some Banach space $\mathcal{B}$. For any
function $B : \R \to \mathcal{B}$, the solution of 
\begin{equation} \label{eq:inhom}
\partial_{t} Y( t) \, = \, A( t) Y( t) \, + \, B( t),
\end{equation}
with boundary condition $Y(0) = y_0$, satisfies the bound
\begin{equation} \label{eq:yest}
\| \, Y( t) \, - \, \gamma_{t}(y_0) \, \| \, \leq \, \int_0^{ t}  \| \, B( t')
\, \| \, d t' .
\end{equation}
\end{lem}

\begin{proof}
For any $t \in \R$, let $X( t)$ be the solution of 
\begin{equation} \label{eq:fode1}
\partial_{t} X( t) \, = \, A( t) \, X( t)
\end{equation}
with boundary condition $X(0) = x_0$, and let $\gamma_{t}$ be the
linear mapping which takes $x_0$ to $X( t)$. By variation of constants,
the solution of the inhomogeneous equation (\ref{eq:inhom}) may be
expressed as
\begin{equation} \label{eq:ysol}
Y( t) \, = \, \gamma_{t} \left( \, y_0 \, + \, \int_0^{
      t} ( \gamma_s)^{-1} \left( B(s) \right) ds \, \right).
\end{equation}
The estimate (\ref{eq:yest}) follows from (\ref{eq:ysol}) as $A( t)$ is
norm preserving.
\end{proof}
\end{appendix}

\subsection*{Acknowledgements}
         
This article is based on work supported by the U.S. National Science
Foundation under Grant \# DMS-0303316. Y.O. is supported by the Japan
Society for the Promotion of Science.
We would like to acknowledge Frank Verstraete for
posing the questions, the answers to which form the basis
of this short note.

\end{document}